\documentclass[conference]{IEEEtran}
\IEEEoverridecommandlockouts

\usepackage{cite}
\usepackage{amsmath,amssymb,amsfonts}
\usepackage{graphicx}
\usepackage{textcomp}
\usepackage{xcolor}
\def\BibTeX{{\rm B\kern-.05em{\sc i\kern-.025em b}\kern-.08em
    T\kern-.1667em\lower.7ex\hbox{E}\kern-.125emX}}

\usepackage{amsmath, amssymb}
\usepackage{mathtools}
\usepackage[small]{caption}
\usepackage{amsthm,multirow,color,amsfonts}
\usepackage[ruled,linesnumbered]{algorithm2e}
\usepackage{tabulary}
\usepackage{subfigure}
\usepackage{graphicx}
\usepackage{setspace}
\usepackage{enumerate}
\usepackage{comment}
\usepackage{multicol,lipsum}
\usepackage{cite}
\usepackage{bm}
\usepackage{bbm}
\usepackage[noend]{algpseudocode}
\usepackage{hyperref}
\usepackage{caption}
\usepackage{subcaption}
\usepackage{textcomp}
\usepackage{xcolor}
\usepackage{subcaption}
\usepackage{booktabs,tabularx}
\makeatletter
\def\BState{\State\hskip-\ALG@thistlm}
\makeatother
	
\usepackage{soul}

\IEEEaftertitletext{\vspace{-1\baselineskip}}
    
\begin{document}

\title{Design Insights into Partition Placement and Routing for DNN Inference in Multi-Hop Edge Networks
}

\author{\IEEEauthorblockN{Jinkun Zhang \,and\, Poonam Yadav}
\IEEEauthorblockA{Department of Computer Science\\
University of York, UK}
}

\maketitle

\begin{abstract}
Partitioned DNN inference is a promising approach for latency-sensitive intelligent services in edge networks, since it allows different parts of a model to be executed across end devices, edge servers, and the cloud. However, in a multi-hop edge network, partition placement and inference traffic routing are inherently coupled: raw inputs, intermediate features, and final outputs may have very different sizes, while candidate nodes also differ in computation capability. In addition, both communication and computation delays can become congestion-dependent under load. In this paper, we study joint partition placement and routing for fixed-partition DNN inference over heterogeneous multi-hop edge networks. We consider a small number of DNN partitions, each placed at exactly one node without replication, and formulate a congestion-aware mixed discrete--continuous optimization problem that captures both routing and execution costs. To solve it, we develop a practical alternating framework that couples partition placement with congestion-aware forwarding updates. Through numerical evaluation on hierarchical, regular, synthetic irregular, and real backbone-inspired topologies, we show that split flexibility is particularly important in IoT--edge--cloud settings, while congestion-aware refinement becomes increasingly beneficial as the offered load grows. We further illustrate how the preferred operating point depends on the communication--computation tradeoff.
\end{abstract}

\section{Introduction}
Deep neural network (DNN) inference is increasingly becoming a core building block for latency-sensitive intelligent services deployed at the network edge, such as real-time perception and interactive mobile AI applications~\cite{teerapittayanon2017distributed}. However, executing an entire DNN locally is often difficult on resource-constrained end devices, while sending raw inputs to a remote cloud can incur substantial communication latency and bandwidth overhead~\cite{kang2017neurosurgeon}, or raise privacy and security considerations~\cite{yang2024penetralium}. These limitations have made partitioned or collaborative DNN inference a compelling alternative, in which different portions of a model are executed across end devices, edge servers, and possibly the cloud so as to better exploit distributed computation and communication resources. 
This paradigm is particularly appealing in collaborative edge environments, where heterogeneous nodes can jointly support DNN inference and model execution naturally interacts with the underlying network.

{ Although partitioned DNN inference has been widely studied, much of the literature still relies on simplified device--edge--cloud pipelines, predetermined collaboration structures, or restrictive communication assumptions~\cite{ren2023survey,kang2017neurosurgeon}. Recent works begin to consider richer distributed settings, including heterogeneous edge cooperation, fine-grained partitioning, and network-aware utility optimization~\cite{li2024distributed,li2024online,ng2024collaborative}. In multi-hop edge networks, however, inference traffic may traverse multiple intermediate links and nodes, making communication an integral part of execution rather than a single offloading step. This is especially important for partitioned DNN inference, since raw inputs, intermediate activations, and final outputs can differ in size, while candidate nodes may also vary in computation capability~\cite{li2024distributed}. Consequently, partition placement and traffic routing are inherently coupled, yet this coupling is often overlooked through fixed pipelines, fixed paths, or formulations that optimize only part of the end-to-end communication--computation process~\cite{cheng2025privacy,ren2023survey}. These observations motivate us to study congestion-aware partition placement and routing for partitioned DNN inference over multi-hop edge networks.}

\begin{figure}
    \centering
    \includegraphics[width=1\linewidth]{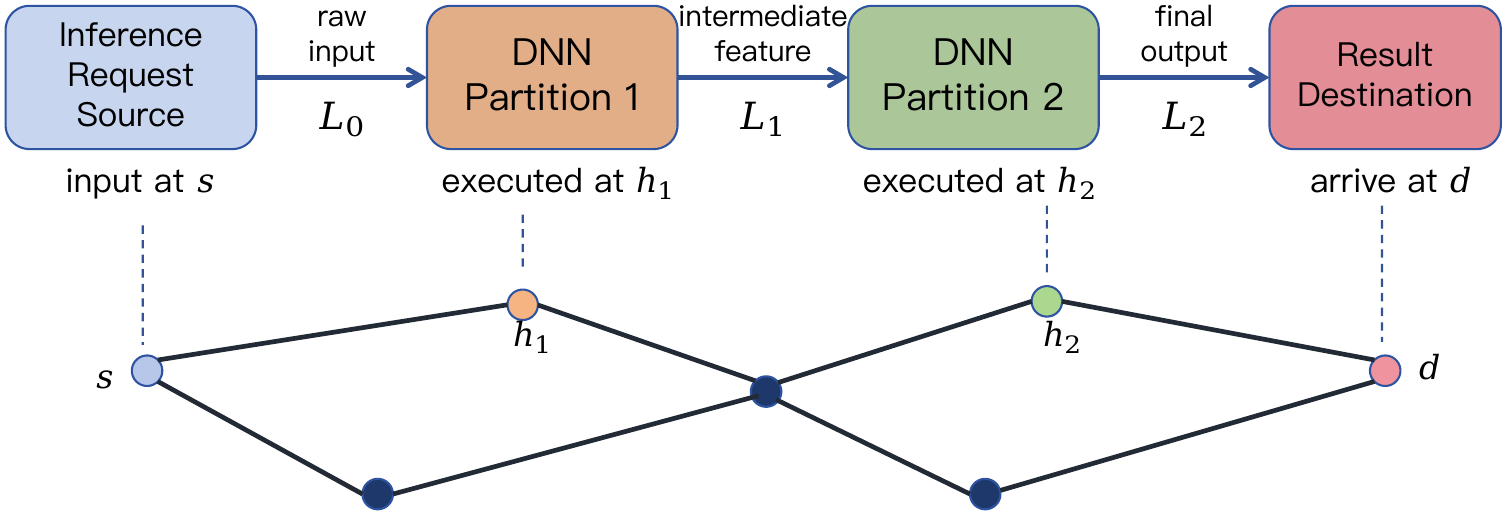}
    \caption{Inference with DNN partitions in a multi-hop network}
    \label{fig:dnn_partition}
\end{figure}

Fig.~\ref{fig:dnn_partition} illustrates the partitioned DNN inference model. The raw input of size $L_0$ originates at source $s$, the two DNN partitions are executed at nodes $h_1$ and $h_2$, and the final output of size $L_2$ is delivered to destination $d$. The intermediate feature of size $L_1$ is transmitted between the two partition locations. Since $h_1$ and $h_2$ may differ and all transmissions may traverse multiple hops, partition placement and routing must be optimized jointly.

A recent work studied delay-optimal forwarding and offloading for generic service-chain applications over arbitrary multi-hop networks with congestion-aware communication and computation costs~\cite{zhang2024delay}. It closely aligns with our motivation, since partitioned DNN inference can also be viewed as a multi-stage computation service. 
However, \cite{zhang2024delay} assumes continuous stage-wise forwarding/offloading, allowing each stage flow to be fractionally routed to or processed by an arbitrarily set of nodes. 
Whereas in practical DNN partitioning, typically only a small number of deployment blocks are involved (in most cases, only $2$), since partitioning and replicating blocks across many nodes can be infeasible due to constraints on communication, memory, power, etc.~\cite{hao2026dnn}.
Here, we focus on a small number of fixed DNN partitions, each placed at exactly one node without replication, which introduces explicit discrete placement decisions coupled with traffic routing. 
Our problem is not a direct specialization of the continuous formulation in \cite{zhang2024delay}, and our focus shifts from theoretical optimality to application-driven design insights.

Our method addresses partitioned DNN inference over a multi-hop edge network with heterogeneous communication and computation resources. We consider a small number of fixed DNN partitions, motivated by the fact that practical deployment typically involves only a few blocks, while fine-grained partitioning and replication across many nodes can be prohibitively expensive in communication, memory, and power. Each partition is placed at exactly one node without replication, and the resulting inference traffic is routed through the network from the source to the selected partition locations and finally to the destination. To capture the interaction between networking and computation, we model both transmission and processing costs as congestion-dependent, so that the placement of DNN partitions and the routing of the induced traffic must be determined jointly. This leads to a mixed discrete--continuous optimization problem, in which partition locations are discrete decisions while forwarding is a continuous traffic allocation variable. Rather than pursuing a globally optimal but computationally prohibitive formulation, we develop a practical congestion-aware solution framework that couples partition placement with routing refinement. Using this framework, we further study how the preferred split execution pattern depends on feature-size variation across partitions, heterogeneity in node computation capability, and the level of network congestion.

Our contributions are summarized as follows:
\begin{itemize}
    \item We formulate a congestion-aware joint partition placement and routing problem for partitioned DNN inference over heterogeneous multi-hop edge networks, where each fixed DNN partition is placed at exactly one node without replication.
    
    \item We develop a practical solution framework for the resulting mixed discrete--continuous problem, by coupling discrete partition placement decisions with congestion-aware forwarding updates.

    \item Through numerical evaluation, we show how the preferred split execution pattern depends on feature-size variation, resource heterogeneity, and offered load, and we quantify the roles of congestion awareness, alternating refinement, and split flexibility.
\end{itemize}

\section{Network formulation}
\label{sec:formulation}

We consider a multi-hop edge network modeled by a directed graph $\mathcal{G}=(\mathcal{V},\mathcal{E})$, where $\mathcal{V}$ and $\mathcal{E}$ denote the sets of nodes and communication links, respectively. A node $i\in\mathcal{V}$ may represent an end device, an edge server, or the cloud, and both nodes and links are allowed to have heterogeneous computation and transmission capabilities. Let $\mathcal{A}$ denote the set of DNN inference services supported by the network. 
For presentation simplicity, we assume each $a \in \mathcal{A}$ uses a DNN with two fixed (vertical) partitions, which are executed sequentially to transform the input data into the final inference result. Each $a$ is associated with a source node $s_a\in\mathcal{V}$, a destination node $d_a\in\mathcal{V}$, and an exogenous input rate $\lambda_a$ (requests/sec)\footnote{ $s_a=d_a$ corresponds to the common case where inference results are sent back to the user generating request. We also allow $s_a \neq d_a$.}. We represent the corresponding inference process by three traffic stages: stage $0$ denotes the raw input data, stage $1$ denotes the intermediate activation generated after the first partition, and stage $2$ denotes the final output generated after the second partition. Let $L_{a,k}$ denote the packet size of stage $k\in\{0,1,2\}$ for application $a$.

For each application $a\in\mathcal{A}$ and partition index $p\in\{1,2\}$, we introduce a binary placement variable $x_i^{a,p}\in\{0,1\}$ for every node $i\in\mathcal{V}$, where $x_i^{a,p}=1$ indicates that partition $p$ of application $a$ is placed at node $i$, and $x_i^{a,p}=0$ otherwise. Since we consider non-replicated partition placement, each partition must be instantiated at exactly one node, i.e.,\footnote{The number ``$1$'' in \eqref{eq:placement_unique} can be changed if allow DNN partition replication.}
\begin{equation}
    \sum_{i\in\mathcal{V}} x_i^{a,p} = 1, \qquad \forall a\in\mathcal{A},\; p\in\{1,2\}.
    \label{eq:placement_unique}
\end{equation}
Under this setting, stage-$0$ traffic is routed toward the host node of partition $1$, where it is processed and converted into stage-$1$ traffic; similarly, stage-$1$ traffic is routed toward the host node of partition $2$, where it is converted into stage-$2$ traffic. Therefore, unlike~\cite{zhang2024delay}, local computation is no longer treated as an independent forwarding option, but dependent on the DNN partition placement scheme. 
Privacy, security, or hardware constraints can be incorporated via feasible host sets $\mathcal{V}_{a,p}\subseteq\mathcal{V}$ with $x_i^{a,p}=0$ for $i\notin\mathcal{V}_{a,p}$, but are omitted here as they are not our focus. 

Inspired by \cite{zhang2024delay}, we use a node-based representation to describe traffic forwarding in the network. 
Let $t_i^{a,k}$ denote the traffic rate (requests/sec) of stage $k\in\{0,1,2\}$ for application $a$ observed at node $i$, including both traffic generated locally at $i$ and traffic forwarded from other nodes. For each link $(i,j)\in\mathcal{E}$, let $\phi_{ij}^{a,k}\in[0,1]$ denote the fraction of stage-$k$ traffic at node $i$ that is forwarded to node $j$. Since local computation is fully determined by the placement variables, the forwarding variables satisfy\footnote{For notational convenience, we let $\phi_{ij}^{a,k}\equiv 0$ whenever $(i,j)\notin\mathcal{E}$.}
\begin{subequations}
\begin{equation}
    \sum_{j\in\mathcal{V}} \phi_{ij}^{a,0} + x_i^{a,1} = 1,\qquad
    \sum_{j\in\mathcal{V}} \phi_{ij}^{a,1} + x_i^{a,2} = 1,
    \label{eq:phi_stage01}
\end{equation}
for all $i\in\mathcal{V}$ and $a\in\mathcal{A}$. 
For the final stage, the traffic exits the network once it reaches the destination, and hence
\begin{equation}
    \sum_{j\in\mathcal{V}} \phi_{ij}^{a,2} =
    \begin{cases}
        0, & i=d_a,\\
        1, & i\neq d_a.
    \end{cases}
    \label{eq:phi_stage2}
\end{equation}
\label{eq:phi_conservation}
\end{subequations}
Eq. \eqref{eq:phi_stage01} implies that a DNN partition host node absorbs and processes the incoming flow at corresponding stage, whereas a non-host node simply forwards it onward;  \eqref{eq:phi_conservation} guarantees that traffic of each stage is routed and terminated in a manner consistent with the DNN execution process: by being processed at the corresponding partition or by exiting the network at the destination.
Since each DNN partition execution represents one request, and will generate one next-stage packet, thus\footnote{$t_i^{a,k}$ can be recursively calculated by \eqref{eq:t_stage}, where $t_i^{a,0} = \lambda_a$ at node $d_a$.}
\begin{equation}
\begin{aligned}
    t_i^{a,0} &= \sum_{j\in\mathcal{V}} t_j^{a,0}\phi_{ji}^{a,0} + \lambda_a \mathbf{1}_{\{i=s_a\}}, 
\\    t_i^{a,1} &= \sum_{j\in\mathcal{V}} t_j^{a,1}\phi_{ji}^{a,1} + x_i^{a,1} t_i^{a,0}, 
\\   t_i^{a,2} &= \sum_{j\in\mathcal{V}} t_j^{a,2}\phi_{ji}^{a,2} + x_i^{a,2} t_i^{a,1}.
\end{aligned}
\label{eq:t_stage}
\end{equation}

We next formulate the communication and computation costs.
For each application $a\in\mathcal{A}$, stage $k\in\{0,1,2\}$, and link $(i,j)\in\mathcal{E}$, let
\begin{equation}
    f_{ij}^{a,k} = t_i^{a,k}\phi_{ij}^{a,k}
\end{equation}
denote the stage-$k$ traffic rate of application $a$ carried on link $(i,j)$. The total traffic load (bit/sec) on $(i,j)$ is then given by
\begin{equation}
    F_{ij} = \sum_{a\in\mathcal{A}}\sum_{k=0}^{2} L_{a,k} f_{ij}^{a,k}.
    \label{eq:link_load}
\end{equation}
Let $w_i^{a,p}$ be the per-request computation workload incurred at node $i$ for processing partition $p\in\{1,2\}$ of application $a$. It may be affected by the DNN partition size, the node's GPU capacity, etc., and can be directly measured from a test run. 
Since partition execution is fully determined by the placement variables, the total computation load at $i$ is
\begin{equation}
    G_i = \sum_{a\in\mathcal{A}}\sum_{p=1}^{2} w_i^{a,p}\, x_i^{a,p} t_i^{a,p-1}.
    \label{eq:comp_load}
\end{equation}
We model the nonlinear transmission cost on link $(i,j)$ by $D_{ij}(F_{ij})$ and the computation cost at node $i$ by $C_i(G_i)$, where both $D_{ij}(\cdot)$ and $C_i(\cdot)$ are assumed to be increasing, continuously differentiable, and convex, with $D_{ij}(0)=0$ and $C_i(0)=0$. Compared to commonly used linear costs, these nonlinear cost functions better capture congestion effects (e.g., queueing) on communication and computation resources. For example, in an M/M/1 queue with service rate $\mu$,
\[
D(F)=\frac{F}{\mu-F}
\]
gives the average number of packets waiting in the queue or being served. Similar convex increasing functions can be used to model congestion at computation nodes hosting DNN partitions. 
They can also mimic the link/GPU capacity constraints~\cite{zhang2024delay}. When both $D(\cdot)$ and $C(\cdot)$ are interpreted as queue lengths, by Little's law, the aggregate cost is proportional to the expected request end-to-end delay.

Let $\bm{x}=[x_i^{a,p}]$ and $\bm{\phi}=[\phi_{ij}^{a,k}]$ denote the global partition placement and forwarding variables.
We minimize the aggregate communication and computation cost in the network, cast as the joint partition placement and routing problem:
\begin{equation}
\begin{aligned}
    \min_{\bm{x},\bm{\phi}} \quad J(\boldsymbol{x},\boldsymbol{\phi}) =  
    & \sum_{(i,j)\in\mathcal{E}} D_{ij}(F_{ij})
    + \sum_{i\in\mathcal{V}} C_i(G_i) \\
    \textrm{subject to} \quad &x_i^{a,p}\in\{0,1\}, \,
    \phi_{ij}^{a,k}\in[0,1],
    \\ &\text{ \eqref{eq:placement_unique}--\eqref{eq:comp_load} hold.}
\end{aligned}
\label{eq:main_problem}
\end{equation}
Note that for networks with $2$-hop or longer paths, $D_{ij}(F_{ij})$ and $C_i(G_i)$ are not necessarily convex with respect to $(\boldsymbol{x}, \boldsymbol{\phi})$. 
Problem \eqref{eq:main_problem} is a mixed integer non-convex optimization problem, where the binary variables $\bm{x}$ determine the partition placement and the continuous variables $\bm{\phi}$ govern stage-wise traffic forwarding over the network.

\section{Solution Approach}


Problem~\eqref{eq:main_problem} is challenging due to the coupling between discrete partition placement and continuous traffic forwarding, together with the nonlinear congestion-dependent communication and computation costs. In particular, binary $\bm{x}$ and continuous $\bm{\phi}$ are tightly coupled, capturing a simple yet fundamental engineering intuition: DNN partitions should be placed at favorable nodes to reduce network cost, but what is favorable itself depends on how stage-wise traffic is routed.
To address this difficulty, we adopt a heuristic yet effective alternating optimization framework, which iteratively updates partition placement and traffic forwarding strategies.

We alternately update partition placement and traffic forwarding strategies. The key idea is that once one set of variables is fixed, the other becomes significantly easier to optimize. In particular, for a given partition placement, the forwarding step updates the stage-wise traffic routing toward the designated partition hosts. Conversely, for a given forwarding solution, the placement step reselects partition hosts to reduce the anticipated global cost under the current network state. The alternating framework is summarized in Algorithm~\ref{alg:alt}.

\vspace{0.2\baselineskip}
\setlength{\textfloatsep}{0pt}
\begin{algorithm}
\SetKwRepeat{DoWhen}{do}{when}
Start with $m=0$, an initial feasible partition placement $\bm{x}^{0}$ and forwarding strategy $\bm{\phi}^{0}$.\\
\DoWhen{$m<M_{\max}$ and algorithm not converged}
{
Update the forwarding variables under fixed partition placement $\bm{x}^{m}$ to obtain $\bm{\phi}^{m+1}$.\\
Update the partition placement under the current forwarding state $\bm{\phi}^{m+1}$ to obtain $\bm{x}^{m+1}$.\\
Set $m \leftarrow m+1$.\\
}
\caption{Alternating Optimization (\texttt{ALT})}
\label{alg:alt}
\end{algorithm}

Specifically, at iteration $m$, the forwarding subproblem with fixed DNN partition placement $\bm{x}^{m}$ is given by
\begin{equation}
\begin{aligned}
    \min_{\bm{\phi}} \quad 
    & \sum_{(i,j)\in\mathcal{E}} D_{ij}(F_{ij}\big|_{\boldsymbol{x}^m}) \\
    \text{subject to} \quad
    & \text{\eqref{eq:phi_conservation} \eqref{eq:t_stage} hold}, \,\phi_{ij}^{a,k}\in[0,1],
\end{aligned}
\label{eq:forwarding_subproblem}
\end{equation}
since when the partition placement is fixed, computation cost $C_i$ are essentially fixed at all nodes $i$.

Then, under current $\bm{\phi}^{m+1}$, the placement subproblem is
\begin{equation}
\begin{aligned}
    \min_{\bm{x}} \quad 
    & J(\bm{x},\bm{\phi}^{m+1}) \\
    \text{subject to} \quad
    & \text{\eqref{eq:placement_unique} holds}, \,x_{i}^{a,p}\in \{0,1\}.
\end{aligned}
\label{eq:placement_subproblem}
\end{equation}
The forwarding and placement subproblems are both still nontrivial to solve exactly at every outer iteration. Therefore, we adopt low-complexity inexact updates for both subproblems.

\subsection{Forwarding subproblem.}
Given a partition placement $\bm{x}^{m}$, each application $a$ essentially induces three ordered source--destination pairs: $(s_a,h_a^{1})$, $(h_a^{1},h_a^{2})$, and $(h_a^{2},d_a)$, where $h_a^{1}$ and $h_a^{2}$ denote the host nodes of partitions $1$ and $2$ under $\bm{x}^{m}$. The forwarding task is therefore to route the three stages of traffic through the multi-hop network in a multi-path manner, so as to reduce the aggregate communication cost.

We adopt a node-based inexact update following Gallager's minimum-delay routing principle~\cite{gallager1977minimum} and its service-chain extension in~\cite{zhang2024delay}. 
The basic idea is that each node gradually reallocates traffic toward outgoing links with smaller system marginal cost.
Here, ``marginal cost'' represents the marginal increase in the overall system latency, consisting of both the immediate cost incurred on the outgoing link and the downstream cost over the remainder of the route.
Specifically, the marginal cost of forwarding stage-$(a,k)$ traffic from $i$ to $j$ can be written as (we omit subscript $\cdot\big|_{\bm{x}^m}$ for simplicity)
\begin{equation}
    \delta_{ij}^{a,k}
    =
    L_{a,k}D_{ij}'\!\left(F_{ij}\right)
    + q_{j}^{a,k},
    \label{eq:delta_forwarding}
\end{equation}
where $q_{j}^{a,k}$ summarizes the downstream marginal cost-to-go under the current forwarding state, and can be recursively calculated from destination nodes toward upstream \cite{zhang2024delay}. Then, each node performs at most $T_{\boldsymbol{\phi}}$ local forwarding updates:
let
\[
\delta_{i,\min}^{a,k} \triangleq \min_{j\in\mathcal{V}} \delta_{ij}^{a,k},
\qquad
j^\star \in \arg\min_{j\in\mathcal{V}} \delta_{ij}^{a,k}.
\]
For each node $i$, application $a$, and stage $k$, we update
\begin{equation}
\phi_{ij}^{a,k}\leftarrow
\Big[
\phi_{ij}^{a,k}
-\alpha_i^{a,k}\big(\delta_{ij}^{a,k}-\delta_{i,\min}^{a,k}\big)
\Big]_+,
\qquad j\neq j^\star,
\end{equation}
and assign the remaining mass to $j^\star$ so that \eqref{eq:phi_conservation} is preserved.
We remark that a node-blocking mechanism is used to prevent the formation of routing loops. We omit further implementation details, including the exact blocking and update scheduling rules. Please see~\cite{gallager1977minimum,zhang2024delay} for a comprehensive explanation.

\subsection{Placement subproblem.}
Given a forwarding state $\bm{\phi}^{m+1}$, the current traffic rates, link loads, and node computation loads are all determined. The placement task is then to update the partition hosts so as to reduce the anticipated global cost under the current network congestion state. To tackle the NP-hard  placement subproblem \eqref{eq:placement_subproblem} efficiently, we adopt a marginal-cost-based inexact reassignment rule.

The basic idea is to evaluate each candidate host node according to the first-order communication and computation cost it would induce under the current forwarding state. 
Denote the marginal transmission and computation costs as (we omit subscript $\cdot\big|_{\bm{\phi}^{m+1}}$ for simplicity)
\begin{equation}
    \ell_{ij}^{a,k} \triangleq L_{a,k}D_{ij}'\!\left(F_{ij}\right), \quad \kappa_i^{a,p} \triangleq w_i^{a,p}C_i'\!\left(G_i\right).
\end{equation}
Then, for any two nodes $u,v\in\mathcal{V}$, let
\begin{equation}
    \Gamma_{uv}^{a,k}
    \triangleq
    \min_{P:u\rightsquigarrow v}\sum\nolimits_{(i,j)\in P} \ell_{ij}^{a,k},
    \label{eq:Gamma_def}
\end{equation}
where $P:u\rightsquigarrow v$ denotes a directed path from $u$ to $v$. Thus, $\Gamma_{uv}^{a,k}$ is the minimum cumulative marginal transmission cost from $u$ to $v$ under the current link marginal weights, and serves as a first-order surrogate for the communication cost change in the placement update.

We then define \emph{candidate score} for placing partition $1$ of application $a$ at node $i$,
\begin{equation}
    S_{a,1}(i)
    \triangleq
    \Gamma_{s_a i}^{a,0}
    +
    \kappa_i^{a,1}
    +
    \Gamma_{i h_a^{2}}^{a,1},
    \label{eq:score_p1}
\end{equation}
where $h_a^{2}$ is the current host of partition $2$. 
Similarly, the candidate score for placing partition $2$ at $i$ is \footnote{We first update partition $1$ while fixing the current host of partition $2$, and then update partition $2$ using the newly selected host of partition $1$.}
\begin{equation}
    S_{a,2}(i)
    \triangleq
    \Gamma_{h_a^{1} i}^{a,1}
    +
    \kappa_i^{a,2}
    +
    \Gamma_{i d_a}^{a,2},
    \label{eq:score_p2}
\end{equation}
where $h_a^{1}$ is the new host of partition $1$. Each score consists of an upstream communication term, a local computation term, and a downstream communication term.
For each $a$, we sequentially update the placement of partitions $1$ and $2$ by selecting the minimum-score host:
\begin{equation}
    h_a^{1} \leftarrow \arg\min_{i\in\mathcal{V}} S_{a,1}(i),
    \quad
    h_a^{2} \leftarrow \arg\min_{i\in\mathcal{V}} S_{a,2}(i).
\end{equation}
The corresponding binary placement variables $\bm{x}^{m+1}$ are then updated accordingly. This yields a low-complexity discrete reassignment step that approximately decreases the placement-side objective under the current forwarding state.

\vspace{0.5\baselineskip}
\noindent\textbf{Discussions.}
The proposed \texttt{ALT} algorithm is an inexact alternating procedure rather than an exact solver for \eqref{eq:main_problem}. Its per-outer-iteration complexity consists of a forwarding-update part and a placement-update part. For the forwarding subproblem, since the number of stages is fixed to three, one local forwarding sweep updates all stage-wise variables by computing the current traffic/load state and the corresponding system marginal costs over the network. Assuming the downstream marginal costs $q_j^{a,k}$ are obtained by one backward recursion as in~\cite{zhang2024delay}, each such sweep scales linearly with the number of applications and links, i.e., $O(|\mathcal{A}||\mathcal{E}|)$; hence, performing $T_{\phi}$ inner forwarding updates incurs complexity $O(T_{\phi}|\mathcal{A}||\mathcal{E}|)$. For the placement subproblem, the candidate scores are built from shortest-path distances under the current marginal link weights. Because the stage-dependent edge weight $L_{a,k}D'_{ij}(F_{ij})$ differs across stages only by a positive scalar factor, the shortest-path structure is determined by the base link weight $D'_{ij}(F_{ij})$. Thus, for each application, it suffices to run at most four single-source shortest-path computations (from $s_a$ and $h_a^1$ on the original graph, and to $h_a^2$ and $d_a$ on the reversed graph), followed by an $O(|\mathcal{V}|)$ scan over candidate nodes. Using Dijkstra's algorithm for nonnegative weights, the placement update therefore costs $O(|\mathcal{A}||\mathcal{E}|\log|\mathcal{V}|+|\mathcal{A}||\mathcal{V}|)$ per outer iteration.

From an implementation perspective, the forwarding update retains the node-based structure of Gallager-style routing and does not require solving a global optimization problem at each iteration. Therefore, it is naturally amenable to scalable implementation: the update can be carried out in a distributed manner across nodes using marginal-cost message passing, or computed in parallel by a central server with knowledge of the current network state. The placement update is written here in a centralized form for clarity, since it requires comparing candidate host scores across nodes. Nevertheless, the same logic can also be implemented in a distributed or hierarchical manner through suitable message exchange and coordination protocols; we leave such protocol design to future work.

Since both subproblems are solved only approximately, we do not claim convergence to a global optimum. Instead, \texttt{ALT} should be viewed as a practical iterative-improvement algorithm, and we terminate it when the objective value changes by less than a prescribed threshold or when the maximum number of outer iterations is reached. Finally, two limiting cases are worth noting. First, if the communication and computation costs are linear, then the marginal quantities become constant and the method reduces to shortest-path-style routing together with a simple path-plus-processing placement rule. Second, if the two DNN partitions are placed at the same node, the inter-partition communication term vanishes, so the model naturally includes the degenerate case where the entire DNN is effectively executed at a single node.

\section{Numerical Evaluation}

We evaluate the proposed alternating framework on four representative network scenarios, namely \emph{IoT}, \emph{Mesh}, \emph{Small-world}, and \emph{GEANT}. 
Specifically, \texttt{IoT} captures a hierarchical IoT--edge--cloud system with strongly heterogeneous communication and computation resources. \texttt{Mesh} is a regular $5\times 5$ mesh. \texttt{Small-world} is a fixed small-world instance, used to capture irregular shortcut-rich connectivity. \texttt{GEANT} is a real backbone-inspired topology based on GEANT, used to assess the method on a realistic irregular graph.

For all scenarios, applications are generated using a fixed random seed, so that the source--destination pairs and arrival rates are reproducible across all algorithms. Unless otherwise specified, the communication stages have sizes $(L_0,L_1,L_2)$. The first partition is lighter than the second one, reflecting the intended split structure in which the first partition may act as a local compression stage while the second partition performs more computation-intensive inference. 
We use M/M/1 queue-like costs on both links and computing nodes, with $\mu$ and $\nu$ denoting link capacity and computing service rates, respectively.
We compare the following methods:
\begin{itemize}
    \item[a.] \textbf{ALT}: the proposed alternating congestion-aware placement and forwarding method. Starting from a feasible structured initialization, it repeatedly updates forwarding and partition placement using congestion-aware marginal costs until convergence. This is the full version of our method, combining congestion-aware modeling with iterative alternating refinement.

    \item[b.] \textbf{OneShot}: a single-pass variant of the proposed framework. It uses the same congestion-aware objective and the same structured initialization as \texttt{ALT}, but performs only one round of placement/forwarding update instead of repeated alternation. Comparing \texttt{ALT} with \texttt{OneShot} isolates the benefit of iterative alternating refinement.

    \item[c.] \textbf{CongUnaware}: a congestion-unaware shortest-extended-path baseline. It constructs the split execution decision by solving a shortest-path problem on an extended graph, where communication and computation are both modeled as linear costs and no congestion feedback is incorporated. Comparing \texttt{ALT} with \texttt{CongUnaware} isolates the benefit of explicit congestion-aware modeling.

    \item[d.] \textbf{CoLocated}: a baseline that enforces colocated execution of the two partitions. It selects a single common execution node for both partitions and then optimizes forwarding under this restriction. Comparing \texttt{ALT} with \texttt{CoLocated} isolates the benefit of split flexibility, i.e., allowing the two partitions to be placed at different nodes.
\end{itemize}

\begin{figure*}[]
\centering
  \begin{minipage}{.38\linewidth}
    \centering
\footnotesize
\begin{tabularx}{\textwidth}{@{\hspace{4pt}}c@{\hspace{4pt}}|@{\hspace{4pt}}c@{\hspace{4pt}}@{\hspace{4pt}}c@{\hspace{4pt}}@{\hspace{4pt}}c@{\hspace{4pt}}@{\hspace{4pt}}c@{\hspace{4pt}}@{\hspace{4pt}}c@{\hspace{4pt}}@{\hspace{4pt}}c@{\hspace{4pt}}}
  \toprule
  Name & $|\mathcal{V}|$ & $|\mathcal{A}|$ & $\bar{\mu}$ & $\bar{\nu}$ & $(L_0,L_1,L_2)$ & $\bar{\lambda}$ \\
  \midrule
  \texttt{IoT}   & 17 & 20 & 8.0  & 8.0 & $(2.0,0.8,0.3)$ & 3.0 \\
  \texttt{Mesh}  & 25 & 30 & 8.0  & 8.0 & $(2.0,0.8,0.3)$ & 3.0 \\
  \texttt{SW}    & 30 & 40 & 10.0  & 10.0 & $(2.0,0.8,0.3)$ & 3.0 \\
  \texttt{GEANT} & 22 & 30 & 10.0 & 10.0 & $(2.0,0.8,0.3)$ & 3.0 \\
  \bottomrule
\end{tabularx}
\vspace{-0.3\baselineskip}
    \captionof{table}
      { Tested scenarios
        \label{tab:scenarios}
      }
  \end{minipage} 
    \begin{minipage}{.47\linewidth}
    \centering
    \vspace{-0\baselineskip}
    \includegraphics[width=0.98\textwidth]{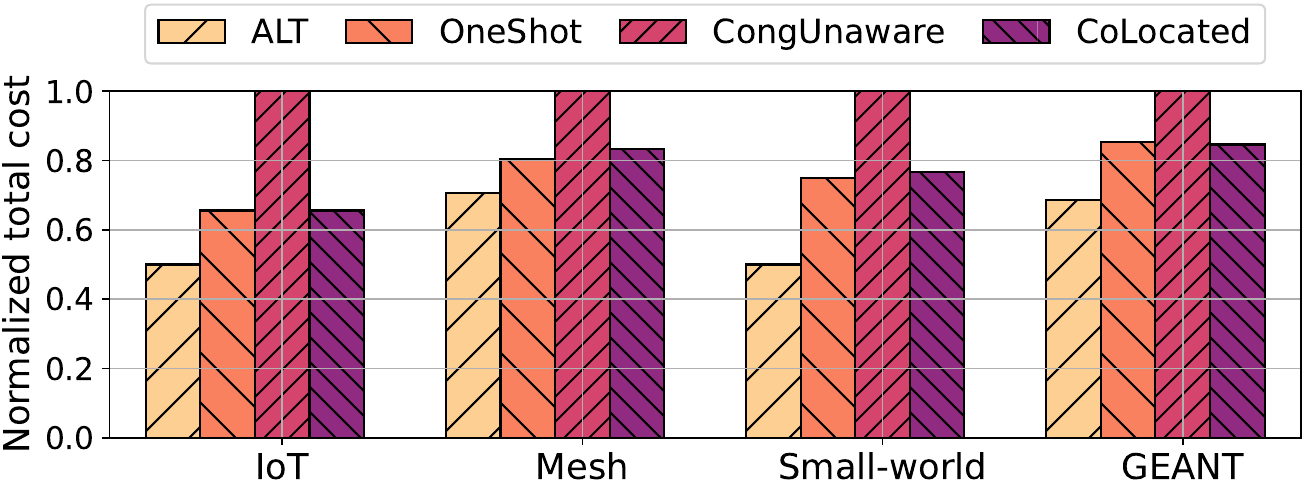}%
    \vspace{-0.1\baselineskip}
    \caption
      {%
        Normalized objective $J$ in all scenarios
        \label{fig:barfig}%
      }%
  \end{minipage}
\vspace{-1\baselineskip}
\end{figure*}

\begin{figure*}[]
     \begin{minipage}{.29\linewidth}
    \centering
    \vspace{-0\baselineskip}
    \includegraphics[width=0.98\textwidth]{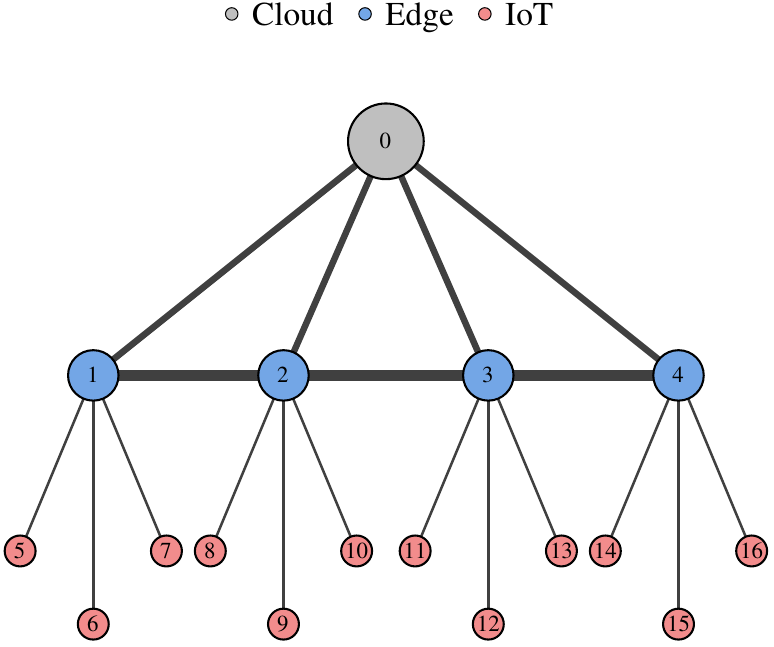}%
    \caption
      {%
        Sample topology (\texttt{IoT})
        \label{fig:sample_network}%
      }%
  \end{minipage}\hfill
    \begin{minipage}{.31\linewidth}
    \centering
    \vspace{-0\baselineskip}
    \includegraphics[width=0.98\textwidth]{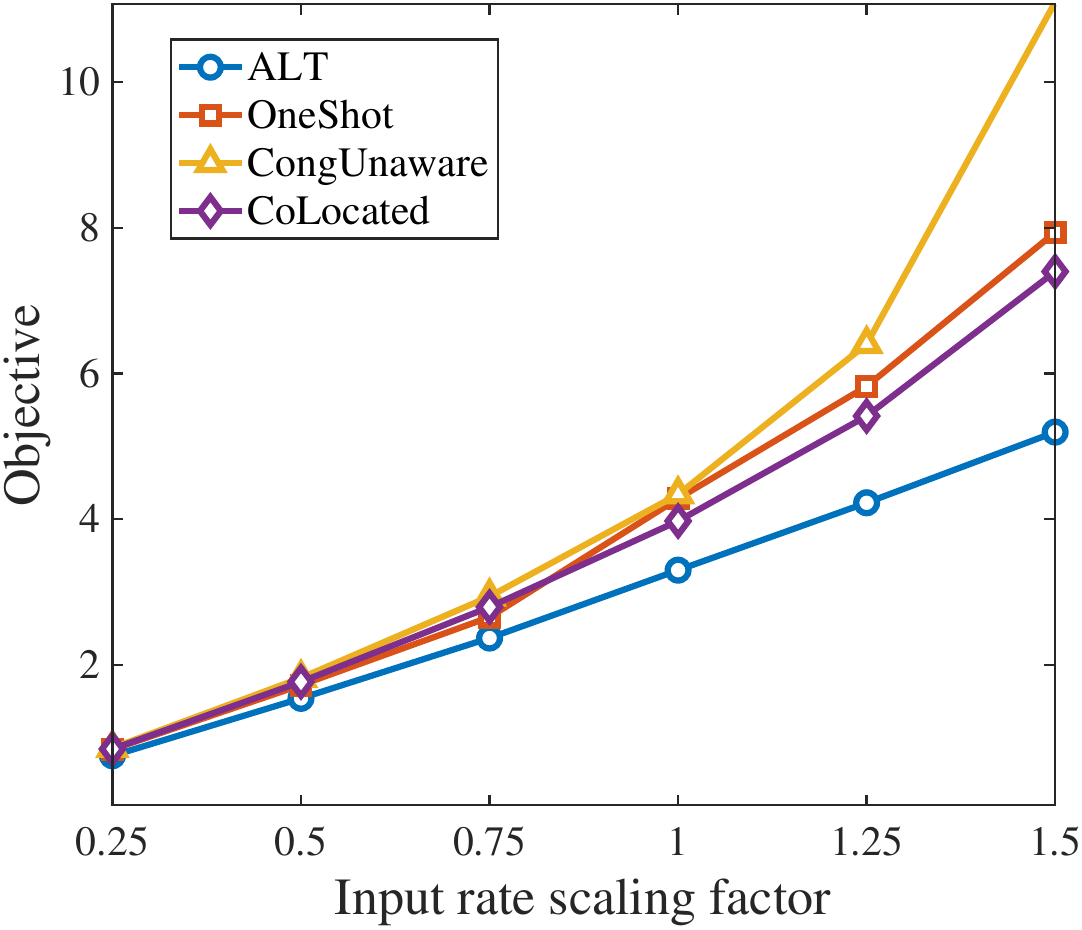}%
    \caption
      {$J$ versus input-rate scaling factor in \texttt{IoT}.
        \label{fig:inputRateScaling}%
      }%
  \end{minipage}\hfill
    \begin{minipage}{.29\linewidth}
    \centering
    \vspace{-0\baselineskip}
    \includegraphics[width=0.98\textwidth]{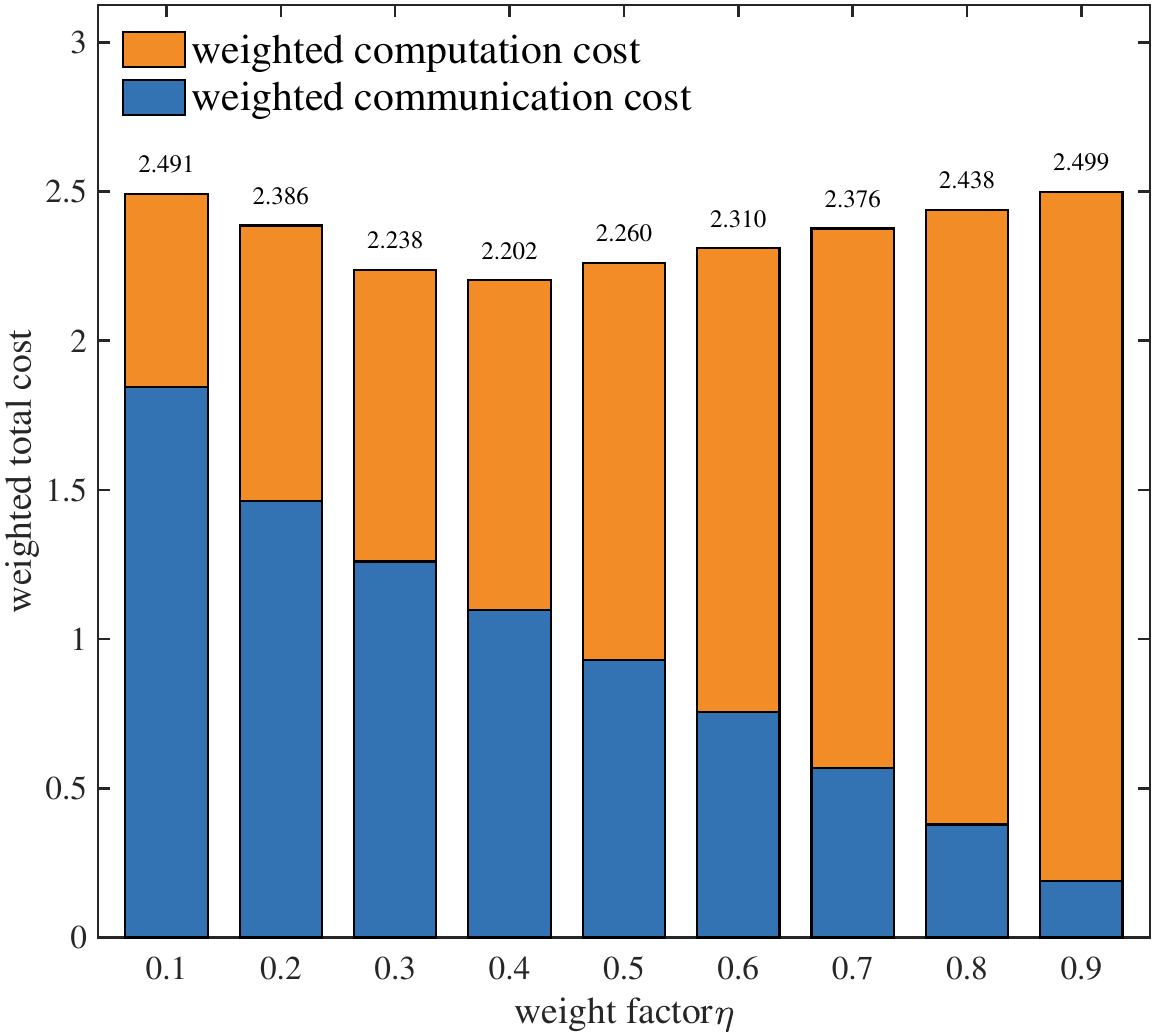}%
    \caption
      {%
       communication--computation tradeoff in \texttt{IoT}
        \label{fig:tradeoff}%
      }%
  \end{minipage}\hfill
\vspace{-1\baselineskip}
\end{figure*}

Table.~\ref{tab:scenarios} summarizes the main experimental settings and the overall comparison, including scenario configurations.
Fig.~\ref{fig:barfig} reports the normalized total cost across all four scenarios. The costs are normalized within each scenario by the largest cost among the compared methods, so that the relative gap is emphasized.

The results reveal three distinct regimes. 
Our proposed method \texttt{ALT} achieves the lowest cost in all tested scenarios.
We observe that, across all scenarios, \texttt{CongUnaware} performs significantly worse than the other methods, indicating that, in congestion-unaware settings, queue-like costs in highly loaded scenarios can incur large expected delays.
In Meanwhile, \texttt{OneShot} is consistently inferior to \texttt{ALT}, since a single placement--forwarding update is generally insufficient to fully capture the mutual dependence between congestion-aware routing and partition placement. Moreover, \texttt{CoLocated} performs poorly because enforcing both partitions to execute at the same node removes the flexibility of split execution, thereby preventing the system from simultaneously exploiting communication-efficient local preprocessing and computation-efficient offloading.

To provide intuition for the hierarchical setting, Fig.~\ref{fig:sample_network} illustrates the IoT topology used in our experiments. The line thickness reflects link bandwidth, the node size reflects computation capability, and the node colors distinguish cloud, edge, and IoT devices. This scenario is intentionally designed to expose the tension between local preprocessing and offloading: IoT devices have weak computation and weak uplinks to edge servers, edge servers are substantially stronger, and the cloud has the highest computation capacity but introduces additional communication cost. This setting explains why split flexibility is particularly valuable in the IoT case.

We next study the effect of increasing workload. Fig.~\ref{fig:inputRateScaling} plots the objective value against a global input-rate scaling factor. The proposed \textsc{ALT} method consistently attains the lowest objective across the tested load range, while the gap relative to \textsc{OneShot}, \textsc{CongUnaware}, and \textsc{CoLocated} widens as the system becomes more heavily loaded. This behavior is precisely the regime in which congestion modeling matters most: under light load, the queueing costs remain close to their linear region, and all methods behave similarly; under higher load, the nonlinear congestion penalties amplify poor routing or poor placement decisions, thereby making repeated congestion-aware refinement increasingly beneficial.

Finally, Fig.~\ref{fig:tradeoff} illustrates the communication--computation tradeoff. We solve the problem under a weighted objective
\[
J_{\eta}=\eta J_{\mathrm{comm}} + (1-\eta)J_{\mathrm{comp}},
\]
where $\eta\in[0,1]$ controls the emphasis on communication cost. The figure reports the weighted total cost together with its communication and computation components. As expected, increasing $\eta$ shifts the optimized solution toward communication-efficient operation, while decreasing $\eta$ prioritizes computation efficiency. Interestingly, the total weighted objective exhibits a shallow minimum at an intermediate value of $\eta$, indicating that neither extreme communication minimization nor extreme computation minimization is universally optimal in this scenario. This observation confirms that the proposed framework is not merely reducing one cost component at the expense of the other; rather, it meaningfully adapts the split placement and forwarding strategy to the desired operating point.


\section{Conclusion} \label{Section:Conclusion}

We studied congestion-aware split execution and forwarding for DNN services in heterogeneous networks with nonlinear communication and computation costs. We proposed \texttt{ALT}, an alternating congestion-aware method that jointly refines placement and forwarding.
The numerical experiment results show that split flexibility is especially important in hierarchical IoT--edge--cloud settings, while congestion-aware refinement becomes more beneficial as the offered load increases. They also show that the preferred solution depends on the communication--computation tradeoff, and that the proposed method can adapt to different operating points.

\section*{ACKNOWLEDGMENT}
This work is supported, in part, by EPSRC and DSIT under grants:
EP/X040518/1, EP/Y037421/1, and EP/Y019229/1.

\bibliographystyle{IEEEtran}
\bibliography{References}

\end{document}